\documentclass[pra,aps,showpacs,twocolumn]{revtex4}
\usepackage{graphicx}
\usepackage{amsmath,amssymb,revsymb}
\usepackage[colorlinks=true,citecolor=blue,linkcolor=blue]{hyperref}

\begin{document}

\title{Universal bound and scattering properties for two dipoles}

\author{Yujun Wang}
\author{Chris H. Greene}
\affiliation{Department of Physics and JILA, University of Colorado, Boulder, Colorado, 80309-0440, USA}

\begin{abstract}
The bound state and low-energy scattering properties of two oriented dipoles are investigated for both bosonic and fermionic symmetries. Interestingly, 
a universal scaling emerges for the expectation value of the angular momentum for deeply-bound two-dipole states. This scaling traces to the 
pendulum motion of two dipoles in strong dipole regime. The low-energy scattering phase shifts of two dipoles also show universal behavior. 
These universal observations make connections to the scaling laws reported in Refs.~\cite{3DBoson,3DFermion} 
for three dipoles. Atomic units are used throughout this work.
\end{abstract}
\maketitle

\section{introduction}
\label{Intro}
The recent experimental realization of ultracold polar molecules~\cite{PolarMol} has attracted tremendous attention through its potential 
future applications in studies of astrophysics, condensed matter physics, quantum computing, and ultracold chemistry~\cite{MolApp1,MolApp2}. 
It also provides great opportunity for studies of few-body physics due to the long-range, anisotropic nature of the molecular interactions in 
the presence of an external electric field.  
For polar molecules that have exothermal reaction paths~\cite{Reactive}, the high short-range reaction probability leads to large ultracold reaction rates 
that have been observed in experiments and have also been explained by a universal 
theory~\cite{UniversalLoss1,UniversalLoss2,UniversalLoss3,UniversalLoss4,UniversalLoss5,UniversalLoss6,UniversalLoss7}.
Novel phenomena involving polar molecules~\cite{MolApp1,MolApp2} are in particular expected from the long-range interactions.  
When the molecules are nonreactive, rich resonant features that are tunable via external field have been predicted in the scattering of two molecules 
in different geometries~\cite{DipoleRes1,DipoleRes2,DipoleRes3,DipoleRes4,DipoleRes5,DipoleRes6,DipoleRes7,DipoleRes8,DipoleRes9,DipoleRes10}.
Interestingly, these resonances are formed with different mechanisms~\cite{DipoleRes4} and therefore occur in an irregular order. 

Whereas two-dipole physics has proven to be surprisingly complicated, the three-dipole physics near two-dipole resonances manifests simple universal behavior for 
both bosonic~\cite{3DBoson} and fermionic~\cite{3DFermion} dipoles. This is quite counterintuitive considering that a three-body problem is generally more 
complicated when compared to a two-body problem. Moreover, in contrast to the previously known three-body systems where 
non-universal short-range physics is important even at unitarity~\cite{BraatenRev}, a system with three dipoles at unitarity can be universally  
defined by two-dipole physics that is characterized by the dipole length $d_\ell$ defined as 
\begin{equation}
d_\ell=m d_m^2/2, 
\end{equation}
where $d_m$ is the magnitude of the electric dipole moment, and $m$ is the mass of a dipole. A close relationship between two- and three-dipole problem is also manifested 
in the effective repulsion between a dipole and a deeply-bound dipolar dimer that has been
identified in Refs.~\cite{3DBoson,3DFermion}.

In the present work we study universal two-dipole physics that can be relevant to the three-dipole physics mentioned above. In particular, 
the properties of deeply- and weakly-bound dipolar states as well as the behavior of low-energy scattering phase shift are discussed for both bosonic and fermionic dipoles. 
It will be shown that when the dipolar interaction is off-resonant, the binding energy of two dipoles scales like $1/m d_\ell^2$, and the size of the corresponding state 
grows linearly with $d_\ell$. For deeply-bound states, the expectation value $\langle \hat{L}^2\rangle$ shows universal $\sqrt{d_\ell}$ scaling, where 
$\hat{L}$ is the angular momentum operator. As will be shown, this scaling traces to the pendulum motion of two dipoles in a strong external field. 
Finally, universal scaling behavior will be shown for both the real part and and the imaginary part 
of the phase shifts that characterize elastic scattering and the scattering into different angular momentum channels, respectively.

This paper is organized as follows. Section~\ref{Theory} introduces our model and method for solving the two-dipole problem. Section~\ref{BoundResult} discusses 
the bound state properties of the dipoles, and Section~\ref{ScattResult} discusses the scattering properties. Finally, we summarize our results in Section~\ref{Summary}.

\section{Theory}
\label{Theory}

To a good approximation at long range the dipolar molecules in an external field can be treated as point dipoles fully-oriented along the field direction. 
The Schr{\"o}dinger equation for two dipoles in spherical coordinates is given by 
\begin{equation}
\left[-\frac{1}{m}\frac{1}{r}\frac{d^2}{dr^2}r+\frac{\hat{L}^2}{m r^2}+V_{\rm dd}\right]\psi=E\psi,
\label{Eq:Schrod}
\end{equation}
where $V_{\rm dd}$ is the interaction between two dipoles that are oriented to the direction of the external field $\hat{z}$:
\begin{equation}
V_{\rm dd}=\frac{2d_\ell}{m}\frac{1-3 (\hat{z}\cdot\hat{r})^2}{r^3}f(r).
\label{Eq:Dipolar}
\end{equation}
The function $f(r)=\tanh(r/r_0)^{16}$ cuts off the dipolar interaction around the short-range length scale $r_0$, 
which avoids an unphysical collapse of the system in the case of the $-1/r^3$ singularity at the origin. 

\subsection{Diabatic representation}
To solve Eq.~(\ref{Eq:Schrod}), it is natural to expand $\psi$ in the basis of angular momentum eigenstates $|l m_l \rangle$ as
\begin{equation}
\psi^{m_l}=\frac{1}{r}\sum_{l'}F_{l'}^{m_l}(r) |l' m_l\rangle,
\end{equation}
where different $m_l$ decouple when the quantization axis is along the direction of the external field.
For large $r$, different $|l m_l \rangle$ states 
decouple and physically serve as scattering channels for two dipoles. In this basis, Eq.~(\ref{Eq:Schrod}) reduces to a set of coupled equations:
\begin{align}
&\left[ -\frac{1}{m}\frac{d^2}{dr^2}+\frac{l(l+1)}{m r^2}\right]F_{l}^{m_l}(r)+\nonumber
\\&\frac{d_\ell}{m r^3}\sum_{l'} D_3(m_l;l,l')  F_{l'}^{m_l}(r)=E F_{l}^{m_l}(r),
\end{align}
where the coupling coefficients $D_3(m_l;l,l')$ are expressed in terms of 3-$j$ symbols:
\begin{align}
D_3(m_l;l,l')=&(-1)^{m_l+1}4\sqrt{(2l+1)(2l'+1)}\nonumber\\
&\left( 
\begin{array}{ccc}
l & 2 & l'\\
0 & 0 & 0
\end{array}
\right)
\left( 
\begin{array}{ccc}
l & 2 & l'\\
-m_l & 0 & m_l
\end{array}
\right).
\end{align}
Note that $l$ and $l'$ are coupled only when $|l'-l|\leqslant 2$, and specifically when $|l'-l|= 2$ for indistinguishable dipoles that are considered in 
the present work. 

\subsection{Adiabatic representation}
Another way to solve Eq.~(\ref{Eq:Schrod}) is to use the adiabatic representation, as has been implemented in Refs.~\cite{DipoleRes4,DipoleRes5,DipoleRes6,DipoleRes7}. 
It has been shown~\cite{DipoleRes4,DipoleRes5,DipoleRes6,DipoleRes7} that the adiabatic representation gives better characterization of 
the various dipolar resonances~\cite{DipoleRes4,DipoleRes6,DipoleRes7}, 
presumably because the adiabatic channels are less coupled than the angular momentum basis at small distances.
At large distances the adiabatic basis reduces to the angular momentum basis, so both representations will give the same scattering matrix. 
Although these two representations are physically equivalent, we would like to compare their efficiencies and use the one that is the most convenient 
for our study that involves both analytical and numerical work.

In the adiabatic representation, the two-dipole wavefunction is expanded in the adiabatic basis $\phi_\nu$ as 
\begin{equation}
\psi^{m_l}=\frac{1}{r}\sum_{\nu'}\tilde{F}_{\nu'}^{m_l}(r)\phi_{\nu'}^{m_l}(r;\Omega),
\label{Eq:AdExp}
\end{equation}
where $\Omega$ represents the polar angle $\theta$ and the azimuthal angle $\varphi$, $\nu$ and $\nu'$ are the adiabatic channel indices.
The adiabatic channel functions $\phi_{\nu'}^{m_l}$ are obtained by solving the adiabatic equation
\begin{equation}
\left(\frac{\hat{L}^2}{mr^2}+V_{\rm dd}\right)\phi_\nu^{m_l}(r;\Omega)=U_\nu(r)\phi_\nu^{m_l}(r;\Omega).
\end{equation}
Upon substitution of the adiabatic expansion Eq.~(\ref{Eq:AdExp}), Eq.~(\ref{Eq:Schrod}) also takes a multi-channel form:
\begin{align}
&\left[ -\frac{1}{m}\frac{d^2}{dr^2}+U_\nu(r)\right]\tilde{F}_{\nu}^{m_l}(r)+\frac{1}{m}\sum_{\nu'}\left[P_{\nu,\nu'}\frac{d}{dr}+\right.\nonumber\\
&\left.\frac{d}{dr}P_{\nu,\nu'}\right]\tilde{F}_{\nu'}^{m_l}(r)+\frac{1}{m}\sum_{\nu'}Q_{\nu,\nu'}\tilde{F}_{\nu'}^{m_l}(r)=E\tilde{F}_{\nu}^{m_l}(r),
\end{align}
but is now characterized by the non-adiabatic couplings $P_{\nu,\nu'}(r)$ and $Q_{\nu,\nu'}(r)$, defined as
\begin{equation}
P_{\nu,\nu'}(r)=\left\langle\!\!\left\langle\phi_\nu\left|\frac{d}{dr}\right|\phi_{\nu'}\right\rangle\!\!\right\rangle, 
Q_{\nu,\nu'}(r)=\left\langle\!\!\left\langle\left.\frac{d}{dr}\phi_\nu\right|\frac{d}{dr}\phi_{\nu'}\right\rangle\!\!\right\rangle.
\end{equation}
The double brackets in the above definition indicate integration over the angles $\Omega$.

In the asymptotic region ($r\gg d_\ell$), the adiabatic channel functions $\phi_\nu$ can be calculated perturbatively from $|l m_l \rangle$, 
which allows us to analytically derive the asymptotic form of the non-adiabatic couplings.
To calculate the leading order of the non-adiabatic couplings $P_{\nu,\nu'}$ for channels $\nu$ and $\nu'$ that asymptotically approach 
the diabatic channels $(l_\nu,m_l)$ and $(l_{\nu'},m_l)$, respectively, the following expansion of the channels functions 
is required: 
\begin{align}
\phi_\nu=\sum_{n=\nu}^{\nu'}\eta_n^{\nu} | l_n m_l \rangle.
\end{align}
The expansion coefficients $\eta_n^\nu$ are obtained by $|n-\nu|$-th order perturbation theory: 
\begin{align}
\eta_n^\nu=C_n^\nu\left(d_\ell/r\right)^{|n-\nu|},
\end{align}
where 
\begin{align}
C_n^\nu=&\prod_{k=0}^{n\!-\!\nu\mp 1}\frac{D_3(m_l;l_\nu\!+\!2k,l_\nu\!+\!2k\pm 2)}{l_\nu(l_\nu\!+\!1)\!-\!(l_\nu\!+\!2k\pm 2)(l_\nu\!+\!2k\pm 2\!+\!1)},\nonumber\\
C_\nu^\nu=&1.
\end{align}
In the above expression the upper sign is taken when $n>\nu$, and the lower sign, $n<\nu$.
The leading order term in the asymptotic expansion of $P_{\nu,\nu'}$ is then calculated as
\begin{align}
P_{\nu,\nu'}\simeq -\frac{d_\ell^{|\nu-\nu'|}}{r^{|\nu-\nu'|+1}} \sum_{n=\nu}^{\nu'\mp 1}|\nu'-n|C_n^\nu C_{n}^{\nu'}. 
\end{align}
Here the upper and lower signs are taken for $\nu'>\nu$ and $\nu'<\nu$, respectively.

Deriving the leading order expression for $Q_{\nu,\nu'}$ requires an expansion of $\phi_\nu$ with two more terms: 
\begin{equation}
\phi_\nu=\sum_{n=\nu\mp 1}^{\nu'\pm 1}\eta_n^{\nu} | l_n m_l \rangle, 
\end{equation}
where the upper and lower signs are taken for $\nu'>\nu$ and $\nu'<\nu$, respectively.
This expansion gives, for $\nu'=\nu$, 
\begin{align}
Q_{\nu,\nu}\simeq \frac{d_\ell^{2}}{r^{4}}\left[(C_{\nu-1}^\nu)^2+(C_{\nu+1}^\nu)^2  \right];
\end{align}
for $|\nu'-\nu|= 1$,
\begin{align}
Q_{\nu,\nu'}\simeq &\frac{d_\ell^{3}}{r^{5}}\left\{2(C_{\nu\pm 1}^\nu C_{\nu'\pm 2}^{\nu'}+C_{\nu\mp 2}^\nu C_{\nu'\mp 1}^{\nu'}) \nonumber \right.\\
&\left.-\left[(C_{\nu\pm 1}^\nu)^2 C_{\nu'\pm 1}^{\nu'}+(C_{\nu'\mp 1}^{\nu'})^2 C_{\nu\mp 1}^\nu \right] \right\};
\end{align}
and for $|\nu'-\nu|\geqslant 2$,
\begin{align}
Q_{\nu,\nu'}\simeq & \frac{d_\ell^{|\nu-\nu'|+2}}{r^{|\nu-\nu'|+4}}\sum_{n=\nu\pm 1}^{\nu'\mp 1} |(n-\nu)(n-\nu')| C_{n\pm 1}^\nu C_{n\mp 1}^{\nu'}.
\end{align}
In all the above expressions for $Q_{\nu,\nu'}$, the upper and lower signs are taken for $\nu'>\nu$ and $\nu'<\nu$, respectively.
All the above asymptotic behavior for the non-adiabatic couplings has been verified by our numerical calculations.

\subsection{Comparison of diabatic and adiabatic representations}
\begin{figure}
\includegraphics[scale=0.31]{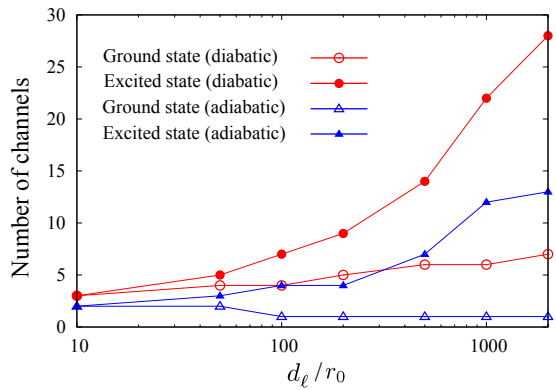}
\caption{
(color online) Number of channels required for converging the dipolar bound states within 1$\%$ in the diabatic (spherical harmonic) and the adiabatic representations. 
}
\label{Fig:Converge}
\end{figure}

In the following we give a brief comparison of the diabatic and adiabatic representations concerning their convergence with number of channels. 
Figure~\ref{Fig:Converge} shows the convergence pattern for the energies of the ground state and the highest excited state with respect to the number of channels as 
$d_\ell$ increases. Generally, the adiabatic representation converges more quickly, and particularly for deeply-lying states it gives better convergence for larger $d_\ell$. 
For highly-lying states more channels are required in order to reach convergence for larger $d_\ell$ in both representations.  

The quick convergence of the adiabatic representation for the deeply-bound dipolar states makes it most advantageous in our analytical study of these states in 
Section~\ref{DeepBoundResult}. 
For studies of the weakly-bound states or scattering, however, 
the adiabatic potentials and non-adiabatic couplings for high-lying channels at small distances ($r\lesssim d_\ell$) 
cannot be analytically determined and require numerical diagonalization. 
Therefore even though the adiabatic representation gives faster convergence it does not provide significantly 
better performance, and we will use the diabatic representation since it can be more easily generated in calculations of weakly-bound states and scattering. 

\section{Bound state properties}
\label{BoundResult}

\subsection{Deeply-bound states}
\label{DeepBoundResult}
In the presence of a strong external field, dipoles in a deeply-bound state tend to align themselves in the linear 
configuration where the dipolar potential Eq.~(\ref{Eq:Dipolar}) is angularly minimal. For small polar angles $\theta$, 
Eq.~(\ref{Eq:Dipolar}) can be approximated by 
\begin{equation}
V_{\rm dd}\approx \frac{2d_\ell}{m}\frac{-2+3\theta^2}{r^3},
\label{Eq:Harmonic}
\end{equation}
which corresponds to a harmonic potential in the $\theta$ direction with trapping frequency of $\omega=\sqrt{24d_\ell/m^2r}$, and 
a zero-point angle given by:
\begin{equation}
\theta_0=\sqrt{\frac{2}{m\omega}}=6^{-1/4}\left(\frac{d_\ell}{r}\right)^{-1/4}.
\end{equation} 

As $d_\ell$ increases, two dipoles in a deeply-bound state become more angularly localized and undergo pendulum motion 
within an angle roughly determined by $\theta_0$. The expectation value of the angular momentum $\langle \hat{L}^2\rangle$ 
is therefore expected to grow with $d_\ell$. 

In view of the connection between $\langle \hat{L}^2\rangle$ and the barrier in 
the collision of a dipole and a dipolar dimer~\cite{3DBoson,3DFermion}, the following studies 
the scaling of $\langle \hat{L}^2\rangle$ with $d_\ell$.
The adiabatic approximation of the two-dipole solution, Eq.~(\ref{Eq:AdExp}), takes the following form: 
\begin{equation}
\psi^{m_l}_{n,\nu}=\frac{1}{r}\tilde{F}_{n,\nu}^{m_l}(r)\phi_{\nu}^{m_l}(r;\Omega),
\end{equation}
where $n$ indicates the approximate quantum number for the radial motion.
The angular wavefunction $\phi_\nu^{m_l}$ for the harmonic potential Eq.~(\ref{Eq:Harmonic}) can be directly written down in terms of 
the associated Laguerre polynomials ${ L}_n^{\alpha}(x)$:
\begin{align}
\phi_\nu^{m_l}=&\sqrt{\frac{2}{\theta_0^2}\frac{\Gamma(\nu+1)}{\Gamma(\nu+|m_l|+1)}}\left(\frac{\theta}{\theta_0}\right)^{|m_l|}
{ L}_\nu^{|m_l|}\!\!\left( \frac{\theta^2}{\theta_0^2}\right) \nonumber\\
&\times e^{-\frac{1}{2}\frac{\theta^2}{\theta_0^2}}\frac{1}{\sqrt{2\pi}}e^{i m_l \varphi}.
\end{align}
By using the approximate angular momentum operator for $\theta\ll 1$:  
\begin{equation}
\hat{L^2}\approx-\frac{1}{\theta}\frac{\partial}{\partial\theta}\theta\frac{\partial}{\partial\theta}-\frac{1}{\theta^2}\frac{\partial^2}{\partial \varphi^2},
\end{equation}
$\langle\hat{L}^2\rangle$ can be determined as:
\begin{equation}
\langle\hat{L}^2\rangle\approx \sqrt{d_\ell}(2\nu+|m_l|+1)\sqrt{6}\left\langle\frac{\tilde{F}_{n,\nu}^{m_l}}{r} \left|\frac{1}{\sqrt{r}}\right| \frac{\tilde{F}_{n,\nu}^{m_l}}{r}\right\rangle.
\label{Eq:AngExp}
\end{equation}
Since for deeply-bound states the radial wavefuction $\tilde{F}_{n,\nu}^{m_l}$ is localized around $r \lesssim r_0$, the radial integral should scale with $1/\sqrt{r_0}$ and 
$\langle\hat{L}^2\rangle$ is expected to obey the following scaling 
\begin{equation}
\langle\hat{L}^2\rangle \propto\sqrt{\frac{d_\ell}{r_0}}(2\nu+|m_l|+1).
\label{Eq:AngExpScale}
\end{equation}
It is interesting to note that this scaling is independent of the identical particle symmetry. This can be simply understood from the angular localization of a pendulum wavefunction 
that makes the exchange effect negligible.

\begin{figure}
\includegraphics[scale=0.38]{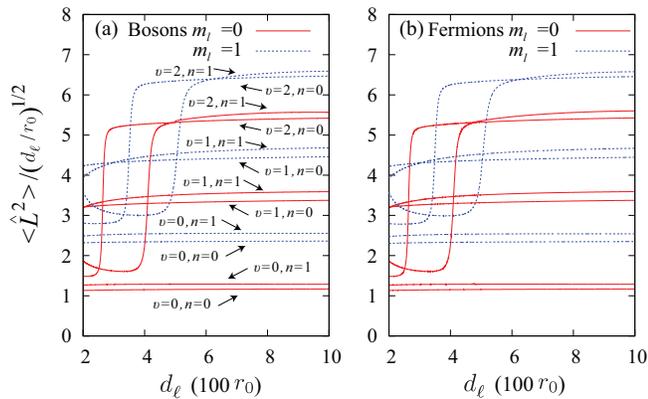}
\caption{
(color online) The scaled angular momentum expectation value $\langle\hat{L}^2\rangle/(d_\ell/r_0)^{1/2}$ for bosonic dipoles (a) and fermionic dipoles (b). 
}
\label{Fig:AngExp}
\end{figure}

\begin{figure}
\includegraphics[scale=0.53]{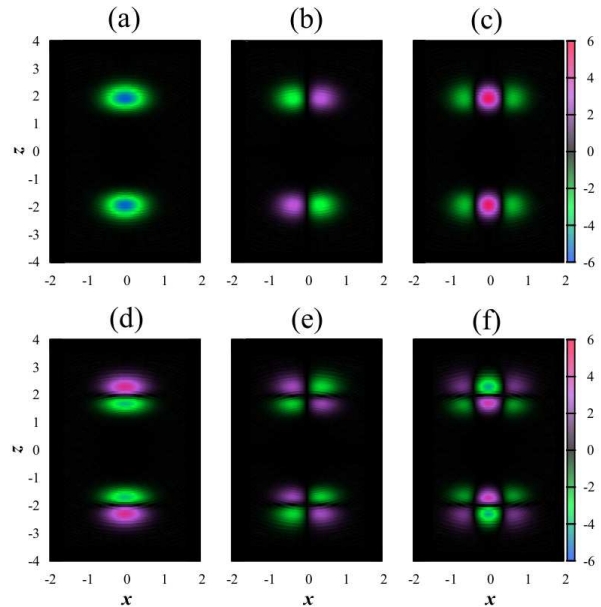}
\caption{
(color online) The cuts of the deeply-bound two-dipole wavefunction along the $x$-$z$ plane for bosons. Different angular and radial excitations $(|m_l|,\nu,n)$ are shown: 
(a) $(0,0,0)$, (b) $(1,0,0)$, (c) $(0,1,0)$, (d) $(0,0,1)$, (e) $(1,0,1)$, 
(f) $(0,1,1)$. The dipole length $d_\ell=600 r_0$ for all these wavefunctions.
}
\label{Fig:AngDensity}
\end{figure}

Figure~\ref{Fig:AngExp} shows the angular momentum expectation value $\langle\hat{L}^2\rangle$ calculated numerically. 
The quick changes in $\langle\hat{L}^2\rangle$ at small $d_\ell$ is related to the transition of the dipolar states from a single angular momentum character to a pendulum character.
The angular and radial excitations for the corresponding 
dipolar states can be seen from the wavefunctions shown in Fig.~\ref{Fig:AngDensity}. 
The equally spaced $\langle\hat{L}^2\rangle$ from numerical calculations suggests the validity of the approximated scaling in Eq.~(\ref{Eq:AngExpScale}). 
The radial excitations $n$ introduce splittings on top of the scaling given by Eq.~(\ref{Eq:AngExpScale}), but such splittings become finer with a sharper short-range cutoff function 
$f(r)$ as both expected from Eq.~(\ref{Eq:AngExp}) and verified by our numerical calculations.

\subsection{Weakly-bound states}
For weakly-bound states, we first discuss the properties of such states away from a dipole-dipole resonance. These properties are important in determining 
the scaling laws for three-dipole recombination~\cite{3DBoson} and possibly also for the vibrational relaxation of a weakly-bound dipolar dimer in an off-resonant situation. 
To this end, we introduce the characteristic size $r_c$ and the characteristic energy $E_c$ for the weakly-bound dipolar states, as the expectation value of $r$ and 
the energy respectively for the state that is right below a zero-energy bound state.

\begin{figure}
\includegraphics[scale=0.35]{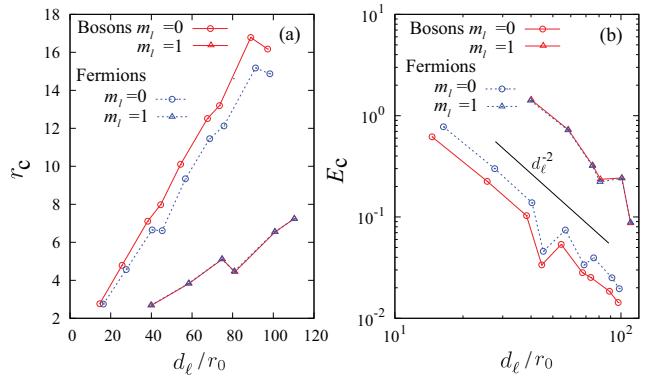}
\caption{
(color online) The characteristic size $r_c$ (a) and the characteristic energy $|E_c|$ (see text) (b) for both bosonic and fermionic dipoles with $|m_l|$$=$0 and 1. 
The irregular jumps come from the avoided crossings between states with different angular momentum character. 
}
\label{Fig:WeaklyBound}
\end{figure}

Figure~\ref{Fig:WeaklyBound} shows numerically calculated $r_c$ and $E_c$ as $d_\ell$ increases. In all cases it is shown 
that $r_c\propto d_\ell$ and $E_c\propto 1/m d_\ell^2$. 
These scalings contrast interestingly with the deeply-bound states, where the sizes of the states shrink to $r_0$ and the 
energies grow deeper linearly with $d_\ell$. 

\begin{figure}
\includegraphics[scale=0.31]{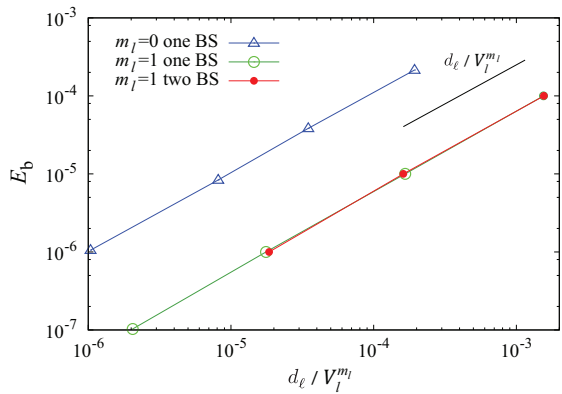}
\caption{
(color online) The scaling behavior of the binding energy $E_b$ with $d_\ell$ and $V_{l}^{m_l}$ near $p$-wave resonances for fermionic dipoles. 
The dipole length $d_\ell$ is tuned to have one or two bound states (BS) 
for $m_l=0$ and $m_l=1$.
}
\label{Fig:Binding}
\end{figure}

Near a dipole-dipole resonance where a zero-energy bound state is formed, 
the binding energy of a weakly-bound state is closely related to the low-energy expansion of the scattering phase shift. In the presence of 
$s$-wave contribution ($m_l$=$0$ for bosons) the binding energy $E_b$ can be written in terms of the $s$-wave scattering length $a_{l=0}^{m_l}$ as 
$E_b\approx 1/m {(a_{l=0}^{m_l})}^2$ by finding the pole in the scattering amplitude~\cite{Landau}. When $s$-wave contribution is absent, searching for a pole in the scattering 
amplitude is not straightforward due to an anomalous low-energy expansion of the phase shifts [see Eq.~(\ref{Eq:PhaseExp})]. Nevertheless, 
near a $p$-wave resonance for fermions we have numerically identified that $E_b\propto d_\ell/V_{l=1}^{m_l}$, where the scattering volume $V_{l}^{m_l}$ diverges near a $p$-wave resonance. 
As shown in Fig.~\ref{Fig:Binding}, the proportionality constant in the scaling of $E_b$ depends on $m_l$ but is independent of 
the short-range interaction details as the number of bound states changes.

\section{Low-energy scattering properties}
\label{ScattResult}
In previous studies of two-dipole elastic scattering~\cite{DipoleRes1,DipoleRes2,DipoleRes3,DipoleRes4,DipoleRes5,DipoleRes6,DipoleRes7,DipoleRes8,DipoleRes9,DipoleRes10}, 
the behavior of the scattering cross-sections have been studied quite extensively. Our goal here, however, 
is to study the two-dipole scattering by examining the low-energy expansion of the phase shifts.
Of particular interest are the phase shifts from the lowest partial wave, which is the most sensitive to the dipole-dipole resonances at zero energy. 

In the present case where the angular momentum is not conserved, we define the phase shifts $\delta_l^{m_l}$ from the diagonal scattering matrix elements $S_{l,l}^{m_l}$:
\begin{equation}
\delta_l^{m_l}=\ln (S_{l,l}^{m_l})/2i.
\end{equation}
In the presence of couplings between different partial waves, $\delta_l^{m_l}$ acquires an imaginary part that characterizes off-diagonal scattering amplitudes. 
The real part of $\delta_l^{m_l}$ controls the elastic scattering.
While dipole-dipole scattering is multichannel, some insights into the low-energy expansion for Re$[\delta_l^{m_l}(k)]$ can be gained from the 
single-channel scattering with a $1/r^3$ potential~\cite{GaoPRA1,SadeghpourReview,GaoPRA2}: 
Re$[\delta_l^{m_l}(k)]\sim -a_l^{m_l} k$, where $k=\sqrt{m E}$ is the scattering wavenumber.

The effective range expansion for short-range potentials~\cite{Bethe,Newton} has the following low-energy expansion of the phase shift:
\begin{equation}
\delta_{l}(k)=-a k -V k^3 + O(k^5),
\end{equation}
where the power of $k$ increases by $2$ for consecutive terms.
For dipolar scattering, however, our numerical study show that the phase shifts are in expansions with increment of $k$ for all partial waves:
\begin{equation}
{\rm Re}[\delta_l^{m_l}(k)]= -a_l^{m_l} k -b_l^{m_l} k^2 -V_l^{m_l} k^3+O(k^4).
\label{Eq:PhaseExp}
\end{equation} 

Our discussion here will be restricted up to the term that starts showing non-universal behavior. 
As has been shown in previous works~\cite{DipoleRes1,DipoleRes2,DipoleRes3,DipoleRes4,DipoleRes5,DipoleRes6,DipoleRes7}, 
the scattering length $a_l^{m_l}$ shows non-universal resonant behavior for $l$=$0$. 
We have further verified that in this case the resonant behavior persists in all higher terms. 
For fermionic dipoles, our numerical study shows non-universal resonant behavior starting from $V_{l=1}^{m_l}$, while $b_{l=1}^{m_l}$ remains universal. 
Non-universal resonant behavior is therefore expected to start from the term of $k^{2l_0+1}$ for only $l$$=$$l_0$, 
where $l_0$ is the lowest partial wave allowed for a given symmetry.
For the terms lower than $k^{2l_0+1}$ the coefficients are expected to be universally determined by $d_\ell$. 

For $l$$>$$0$, Ref.~\cite{DipoleRes5} has analytically derived a universal expression for the $T$-matrix to the leading order in $k$, 
their results can be readily used to give the following expression for $a_l^{m_l}$, 
\begin{equation}
a_l^{m_l}=d_\ell\frac{D_3(m_l;l,l)}{2l(l+1)}.
\end{equation}

\begin{figure}
\includegraphics[scale=0.34]{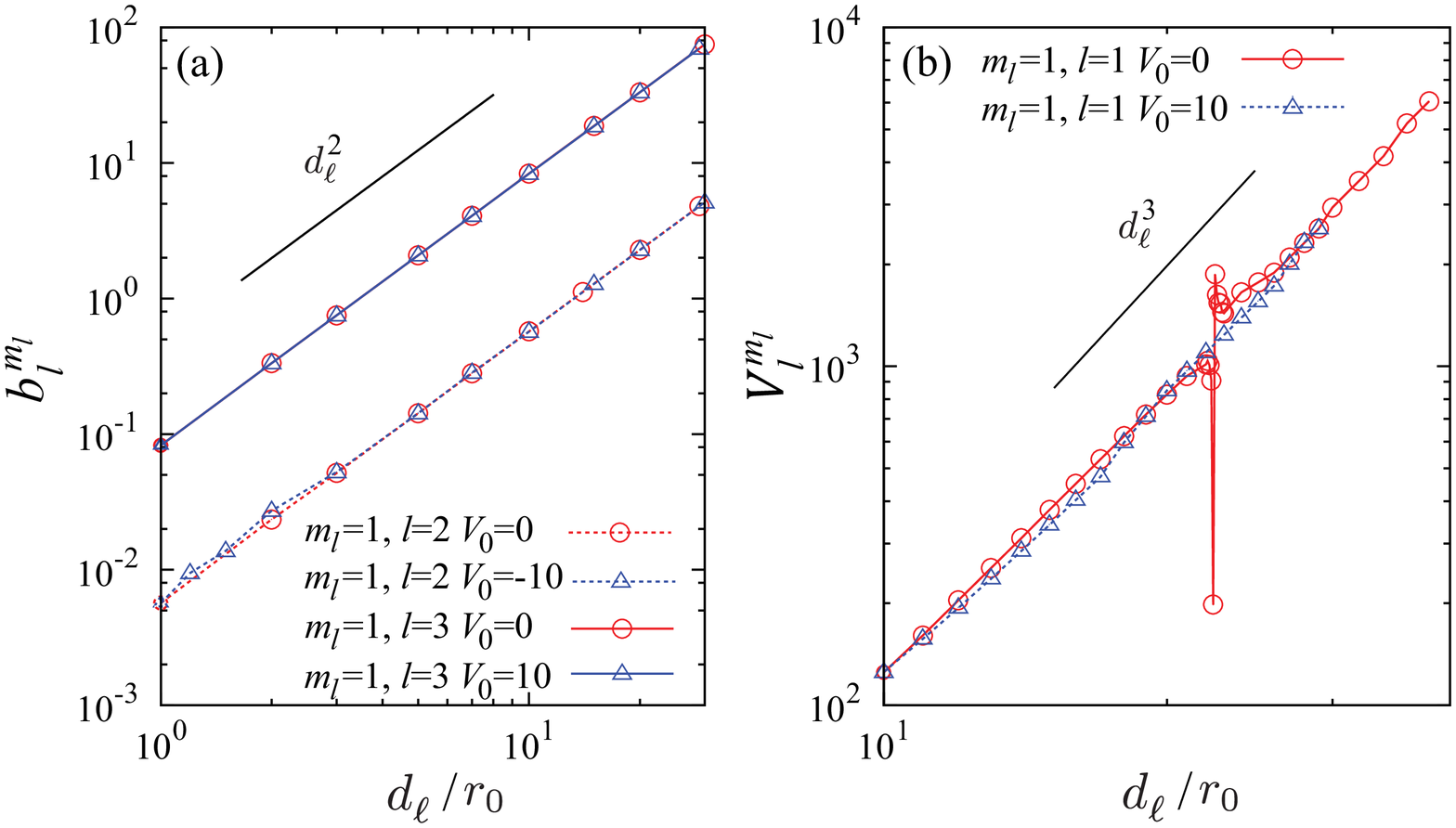}
\caption{
(color online) (a) The $d_\ell$-dependence of the coefficient $b_l^{m_l}$ in the low-energy expansion of Re$[\delta_l^{m_l}(k)]$.
(b) The $d_\ell$-dependence of the scattering volume $V_l^{m_l}$.
}
\label{Fig:Bl_Vl}
\end{figure}

Next we study the scaling behavior of $b_l^{m_l}$ and the scattering 
volume $V_l^{m_l}$ by numerical calculations. 
The universality is tested by adding an isotropic short-range interaction $V_{\rm iso}=V_0{\rm sech}^2(r/r_0)$.
Figure~\ref{Fig:Bl_Vl} (a) shows the $d_\ell$ dependence of a few $b_l^{m_l}$ with different $V_0$. 
It is clearly seen that $b_l^{m_l}$ follows a $d_\ell^2$ scaling behavior that is independent of $V_0$. 
An exception for this universal behavior is $b_{l=0}^{m_l=0}$, where the non-universal resonant behavior already begins in the lower term $a_{l=0}^{m_l=0}$. 
Nevertheless, $b_{l=0}^{m_l=0}$ is found to follow a universal $d_\ell^2$ background scaling with non-universal resonant features on top. 
Table~\ref{Tab:Bl_Vl} lists some numerically determined scaling coefficients.
\begin{table}
\begin{ruledtabular}
\begin{tabular}{ccccc}
&$m_l$ & $l$ &  $b_l^{m_l}$ $(d_\ell^2)$& \\
\hline

&0 & 1 & -2.60$\times$$10^{-1}$& \\ 
&0 & 2 & 2.56$\times$$10^{-2}$&\\
&0 & 3 & 2.30$\times$$10^{-3}$&\\

&1 & 1 & -8.32$\times$$10^{-2}$& \\
&1 & 2 & -5.70$\times$$10^{-3}$&\\
&1 & 3 & 1.27$\times$$10^{-3}$&
\end{tabular}
\end{ruledtabular}
\caption{The numerically calculated universal scaling for the phase shift expansion coefficient $b_l^{m_l}$ for a few symmetries.}
\label{Tab:Bl_Vl}
\end{table}

The study of scattering volume is more challenging due to the difficulty to numerically fit Eq.~(\ref{Eq:PhaseExp}) to the third order accurately. 
Our numerical study shows that the scattering volume $V_l^{|m_l|}$ follows a $d_\ell^3$ scaling in general. 
For $l$$=$1, the lowest partial wave allowed for fermionic scattering, 
non-universal resonant features are expected for $V_{l=1}^{|m_l|}$. Nevertheless, a universal $d_\ell^3$ background scaling can be identified from Fig.~\ref{Fig:Bl_Vl} (b). 
The positions of the resonant features clearly depend on the short-range interaction as tuned by $V_0$, but the background scaling remains unaltered.

Finally we discuss the low-energy expansion for the imaginary part for the phase shift ${\rm Im}[\delta_l^{m_l}(k)]$. The following expansion 
\begin{equation}
{\rm Im}[\delta_l^{m_l}(k)]= -c_l^{m_l} k^2 +O(k^3)
\label{Eq:PhaseExp}
\end{equation} 
is found from our numerical calculations, indicating a vanishing imaginary part in the scattering length when $k d_\ell\ll 1$. 
By using the unitary constraint on the diagonal element of the $S$-matrix to leading order, $c_l^{m_l}$ can be determined as
\begin{align}
c_l^{m_l}\!\!=\!\!-\frac{d_\ell^2}{12}\left\{\left[\frac{D_3(m_l;l,l-2)}{l(l+1)}\right]^2\!\!+\!\!\left[\frac{D_3(m_l;l,l+2)}{(l+1)(l+2)}\right]^2\right\}
\end{align}
for all partial waves.

\section{Summary}
\label{Summary}
To summarize, we have studied the universal properties for two dipoles. The long-range, anisotropic dipolar interaction brings rich, universal physics that  
has key implications for the universal three-dipole physics. 
This is shown particularly for both the deeply-bound and weakly-bound sides of the dipolar spectrum. For the deeply-bound dipolar states, the pendulum motion between the 
dipoles gives rise to a universal growth in the expectation value of the angular momentum, which produces a centrifugal barrier between a dipole 
and a dipolar dimer~\cite{3DBoson,3DFermion}. 
For the weakly-bound states, general scalings of the binding energy and the size of 
the states are identified, despite the complicated level crossings for states with different angular momentum characters. Finally, the low-energy scattering phase shifts 
for two dipoles are predominantly determined by the dipole length, with some non-universal ingredients that give rise to resonant features.
\begin{acknowledgments}
This work is supported in part by the AFOSR-MURI and by the National Science Foundation. We thank J. P. D'Incao and J. L. Bohn for stimulating discussions.
\end{acknowledgments}

\end{document}